To appear in *Deciphering the Ancient Universe with GRBs*, AIPC (N. Kawai and S. Nagataki, eds.), in press (2010)

# GRB Probes of the Early Universe with *EXIST*

Jonathan E. Grindlay[a] and the *EXIST* Team[b]

[a]*Harvard-Smithsonian CfA, 60 Garden St., Cambridge, MA,* [b]*http://exist.gsfc.nasa.gov*

**Abstract.** With the *Swift* detection of GRB090423 at z = 8.2, it was confirmed that GRBs are now detectable at (significantly) larger redshifts than AGN, and so can indeed be used as probes of the Early Universe. The proposed Energetic X-ray Imaging Survey Telescope (*EXIST*) mission has been designed to detect and promptly measure redshifts and both soft X-ray (0.1 - 10 keV) and simultaneous nUV-nIR (0.3 - 2.3microns) imaging and spectra for GRBs out to redshifts z ~18, which encompasses (or even exceeds) current estimates for Pop III stars that are expected to be massive and possibly GRB sources. Scaling from *Swift* for the ~10X greater sensitivity of *EXIST*, more than 100 GRBs at z ≥8 may be detected and would provide direct constraints on the formation and evolution of the first stars and galaxies. For GRBs at redshifts z ≥ 8, with Lyman breaks at greater than 1.12microns, spectra at resolution R = 30 or R = 3000 for afterglows with AB magnitudes brighter than 24 or 20 (respectively) within ~3000sec of trigger will directly probe the Epoch of Reionization, formation of galaxies, and cosmic star formation rate. The proposed *EXIST* mission can probe these questions, and many others, given its unparalleled combination of sensitivity and spatial-spectral-temporal coverage and resolution. Here we provide an overview of the key science objectives for GRBs as probes of the early Universe and of extreme physics, and the mission plan and technical readiness to bring this to *EXIST*.

## INTRODUCTION

With the discovery of GRB090423 at $z = 8.2$ (Salvaterra et al 2009, Tanvir et al 2009), Gamma-Ray Bursts (GRBs) have rapidly surpassed luminous quasars as the most distant objects with spectroscopic redshifts. As discussed by Bromm and Loeb (2007, and references therein), the non-thermal power law afterglow emission (fading with power law decay typically as $t^{-1.3}$) is detectable in its decay earlier (and thus brighter) due to the redshift time dilation factor $1 + z$. Thus GRBs at high z are the optimum probes of the Early Universe provided that spectra of their luminous afterglows can be obtained promptly at wavelengths longward of their spectral cutoffs at redshifted Ly-α, or at ≥1.216μm[$(1 + z)/10$]. A large (>2000) sample of GRBs and redshifts, all with spectral energy distributions (SEDs: nIR-optical-nUV-X-rays-soft γ-rays) and ≥1000 with high resolution (R = 3000) spectra, will enable definitive studies of the Epoch of Reionization (EoR) as well as enable the use of GRBs as probes of the cosmic star formation rate (SFR) and galaxy formation vs. z. Constraints on GRB jets and emission (both long and short) may enable their use as quasi-standard candles and thus probes of cosmic expansion history at redshifts beyond those traced by SNIa's.

In this paper, we first update our previous descriptions (Grindlay et al 2009, 2010) of the proposed *EXIST* mission. We then describe its primary science objective -- to explore the early Universe ($z$ >7) with *GRBs* with its unique wide-field hard X-ray (5-600 keV) survey and immediate (<200sec) followup with sensitive broad-band (IR-visible through soft γ-ray) high resolution imaging and spectra. GRBs are compared with LAE and J-band dropout surveys for studies of the EoR.The initial 2y all-sky scanning survey and then 3y pointed study phase will both produce large samples of GRBs and *Blazars and obscured AGN* (our second mission science goal), to constrain their physics and evolution. These also enable our third key science objective: to discover and study *Transients*, from stellar flares and BHs within the Galaxy, to tidal disruption flares by quiescent SMBHs out to ~200Mpc. Finally, we update the mission development and our recent successful first balloon flight test of *ProtoEXIST1*, which demonstrate that *EXIST* could be launched in this decade.

## OVERVIEW OF *EXIST*

The Energetic X-ray Imaging Survey Telescope (*EXIST*) has been developed since its initial selection for study in 1994 and evolved into an ever more powerful, yet less massive, design (see Grindlay et al 2010). *EXIST* carried out an Astrophysics Strategic Mission Concept (ASMC) Study, in preparation for the 2010 Decadal Survey (Astro2010), and was radically re-designed to be much smaller, less massive and expensive, and fully autonomous for rapid on-board source identifications and redshifts.

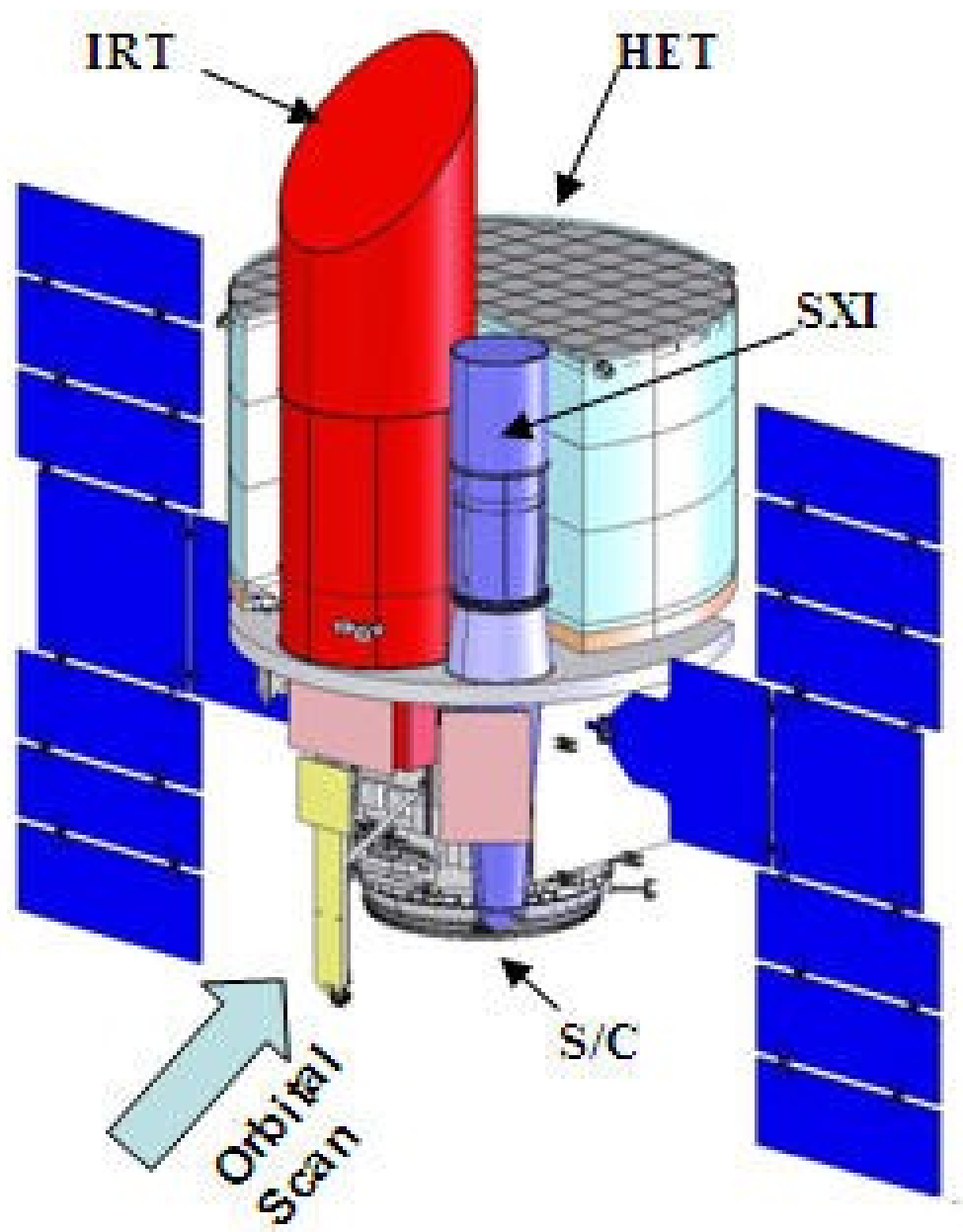

**FIGURE 1.** Wide-field HET (90° x 70°), and narrow-field IRT (5' x 5') and SXI (20') co-aligned on the *EXIST* spacecraft. Zenith pointing ~23° North and South of the $i$ = 15° orbital plane on alternate orbits surveys the full sky every ~3h for HET and half-sky every 6mo for SXI during the 2y scanning survey. IRT operates in pointings (~70% of full mission).

Fig. 1 shows the overall mission payload, with a single large area (4.5 m$^2$) coded aperture High Energy Telescope (HET: 5-600keV) with a hybrid (2-scale) coded aperture mask enabling fast imaging with ~10X finer resolution than *Swift*/BAT; a nUV-optical-nIR (0.3 – 2.3μm) 1.1m aperture telescope (IRT) for simultaneous 4-band imaging or spectroscopy; and a Soft X-ray Imager (SXI) telescope with 0.1 – 10 keV imaging and fast timing. All 3 instruments have significantly finer resolution, broader bands and ≥10*X* the sensitivity of their counterparts (BAT, UVOT and XRT) on *Swift*. The overall instrument parameters for all three instruments are given in Grindlay et al (2009). Details are given for the HET instrument by Hong et al (2009, 2010) and for HET imaging by Skinner et al (2009); for the IRT by Kutyrev et al (2009, 2010) and for the SXI by Tagliaferri et al (2009).

The optimized *EXIST* design (Fig. 1) is conceptually different from *Swift*:

*Scanning vs. pointing:* The orbital scans (Fig. 1) enable both higher sensitivity (by averaging out systematics) and sky coverage (full sky every 3h) for the 2y scanning survey. Followup pointings for GRBs and Transients during the 3y pointed-mode mission phase enables ~full-sky HET coverage each day. Fast readout in the SXI enables photon counting and half-sky surveys every 6months in scanning mode with flux sensitivity ~3 x $10^{-14}$ erg $cm^{-2}$ $s^{-1}$ or ~10X more sensitive than *ROSAT* and with ~4X finer angular resolution (20"). The readout mode of the H2RG detectors and fine guiding for the IRT restrict it to pointing modes (which is ~70% of the 5y mission).

*Deep optical-NIR imaging and spectra:* The 1.1m aperture for the IRT combined with its passively cooled (-30C) optics gives thermal backgrounds at 2 µm that are ~1000X lower than on ground, and free of the bright OH emission lines (Barton et al 2004) that greatly limit ground-based searches. At J, H and K, the IRT is ≥10X faster than Keck!

*Real-time identifications and redshifts:* With initial HET detection (>5σ) positions of <20" (90% conf.), and re-pointings <150s (from trigger) for initial SXI positions of <2", the IRT imaging (0.15" pixels) in 4 bands simultaneously (via 3 dichroics; see Table 1 and Kutyrev et al 2009), GRB IDs will usually be obvious from comparison with 1' deep images generated on board from stored catalogs. The object is moved autonomously onto the low-res (R = 30) objective prism slit for a redshift measurement (AB = 23 in 300s) or directly onto the high-res (R = 3000) long slit (4") for spectra (AB = 19 in 2ksec) to resolve the damped Ly-α absorption (DLA) red wing to measure the ionization of the host galaxy vs. IGM. This will map the epoch of reionization (McQuinn et al 2009).out to $z$ ~ 18 (for sufficiently bright/long GRBs). We note that for z >8 GRBs, time dilation will allow even ~15ksec spectra integrations so that the imaging and spectroscopy limits in Table 1 are ~2mag deeper.

**TABLE 1.** *EXIST*/IRT configuration, modes and sensitivities

| | | | |
|---|---|---|---|
| **Telescope** | 1.1m aperture Cass.; passively cooled -30C; fixed 5' focal plane; 0.15"pix image/spectra | | |
| **Spectral Bands** 2 VIS + 2 IR | 0.3 – 0.52, 0.52 – 0.9µm (*HyViSi*) 0.9 – 1.38, 1.38 – 2.3µm (*H2RG*) | | |
| **Mode** | **Field view (arcmin)** | **Spectral Res.** | **AB @ S/N=10 (req. exp. sec)** |
| Imaging (4 bands) | 3.75' x 4.25' | 3-4 | Vis: 24 (100s) IR: 24 (150s) |
| Low Res. Obj. prism | 3.75' x 0.75' | 30 | Vis: 22 (300s) IR: 22 (300s) |
| Low Res. Single Slit | 20" long slit | 30 | Vis: 23 (300s) IR: 23 (300s) |
| High Res. Single Slit | 4" long slit | 3000 | Vis: 19 (1500s) IR: 19 (2000s) |

*All-timescales, all-time-all-sky imaging:* Because it needs ~100X less telemetry and is also required for scanning, *EXIST* will bring down photon data for HET (and SXI) rather than binned detector images. HET and SXI event data will include all event times (to ~10μsec) as well as energies, position and status so that imaging and spectra can be done on any timescale and energy binning, for any source or sources in any region of sky covered in scanning or pointing. Thus, for example, pulse profiles of X-ray pulsars or lightcurves of AGN can be retrieved over the full mission timescale.

<u>Serendipitous surveys from ~3y pointing mode:</u> Finally, during the 3y pointed mission phase for followup of ~30,000 sources discovered in the 2y scanning survey, and continued slew triggers on GRBs, the SXI will cover ~40% of the sky to sensitivity ~1 x $10^{-14}$ cgs and so conduct a *wide*-field, survey for galaxy clusters, obscured AGN and other objects. With exposures ≥2000sec/field, and simultaneous nUV-nIR high resolution (0.15” pixels) imaging (to AB = 27!) and objective prism spectroscopy partially covered by the IRT (see Table 1) for the central 5’ x 5’ of each SXI field, *EXIST* discoveries would also enable/trigger followup studies with very large telescopes (GMT, TMT) and especially (for z ≥7 host galaxies) with JWST.

## GRB MEASURES OF THE EARLY UNIVERSE TO *EXIST*

Measuring the high-z Universe with GRBs requires prompt (≤300s) GRB identifications from SXI/IRT imaging from which variability or photo-z’s will reveal the source, followed by nIR (and continued soft X-ray) spectroscopy.

### Redshifts for most GRBs and EoR constraints for many

Redshifts and spectra (R = 3000) of a large GRB sample at z >6 are the highest priority. Fig. 2 shows a sample of *Swift* GRBs redshifted to z = 12 (see Grindlay et al 2010) now plotted with the imaging (R = 3) and spectroscopy (R = 30, 3000) sensitivity limits as given in Table 2. It is clear that *all* optical (or nIR) identified GRBs from *Swift* would have redshifts, with R = 30 and most with R = 3000 spectra. Given the deep nIR imaging sensitivity, a significant fraction of dark GRBs would be identified, even for log(NH) ~22, or $A_K$ <2. The key requirement is low background, high sensitivity in the ~1-2.2μm band for red wing measures of Ly-α at z ~7-18 from which to measure the ionization in the host vs. local IGM and constrain the epoch of reionization (EoR), SFR and metallicity of the Universe vs. *z* (McQuinn et al 2009). As shown in Fig. 3, this is generally *not* possible from the ground due to thermal and OH emission line backgrounds and atmospheric transparency. The limits of sensitivity reached by Keck or VLT are indicated by the low resolution spectrum of GRB090423, with J, H ~21 at 17h after the burst (Tanvir et al 2009), that could measure z but not the EoR. In contrast, the ~70% of *Swift* GRBs with optical/nIR AB ≤19-20 could have high resolution (R = 3000) spectra measured at the expected IRT sensitivity limit for a 2000s integration (Kutyrev et al 2009). Each would constrain the local EoR and approximate host galaxy metallicity from both absorption and emission lines (e.g. HeII vs. MgII). GRBs even ~4mag fainter would still have their redshifts measured with the low resolution (R =30) slit or objective prism in 300sec after acquisition in the IRT (or within ≤450s after trigger).

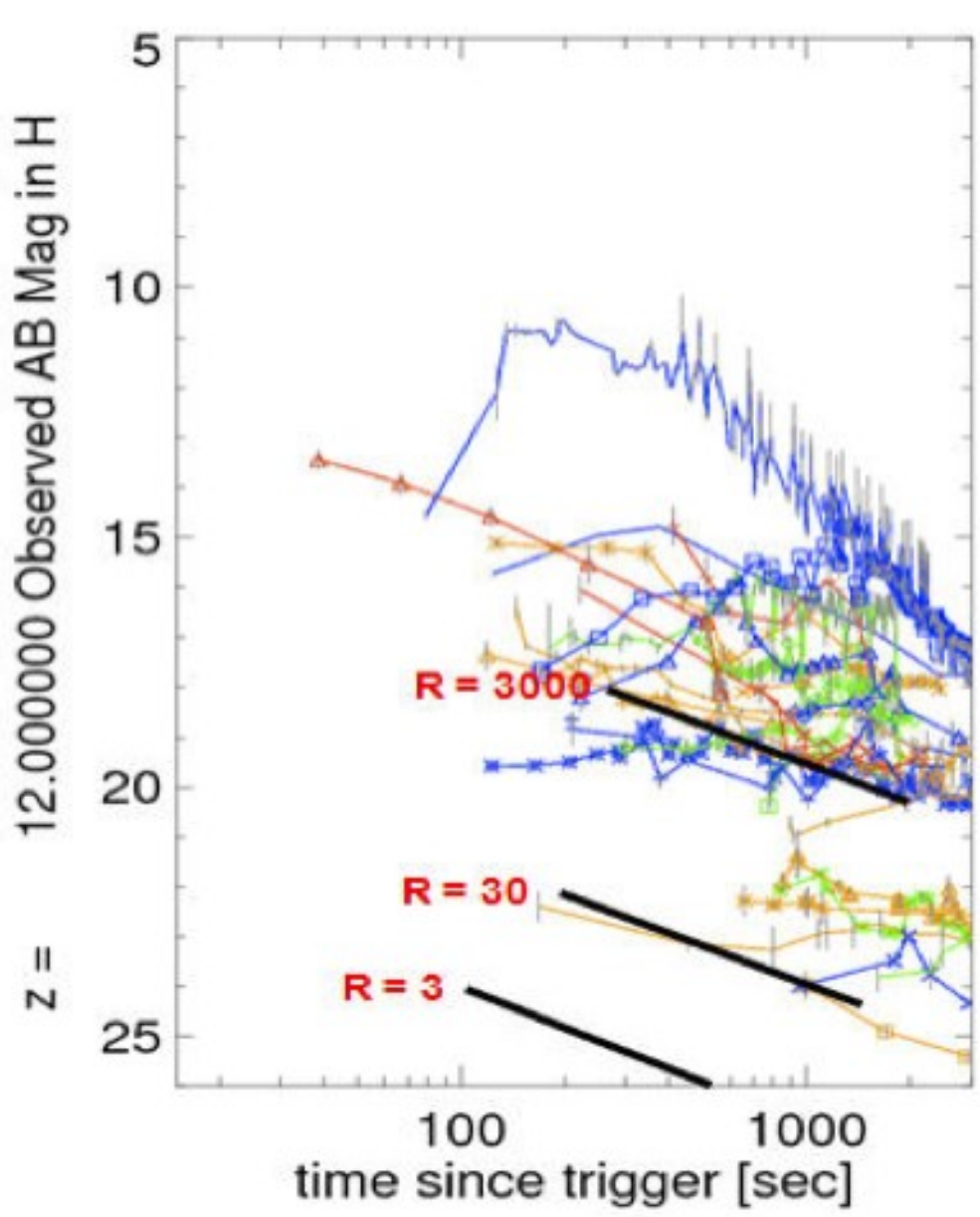

**FIGURE 2.** *Swift* GRB afterglows extrapolated to H band and redshifted to *z* = 12, with sensitivity limits for imaging (R =3) and spectra (R = 30 and 3000) for integration times shown.

*EXIST* is ~7-10X more sensitive to GRBs (particularly high *z*) than *Swift* due to its scanning larger sky coverage for long GRBs, lower energy threshold (5 vs. 15 keV) and lower backgrounds from its active rear shield, which both vetoes Earth albedo background and extends GRB spectra up to ~3MeV. The expected GRB rate is ~400-600/y with a fractional redshift distribution (vs. *Swift*) shown in Fig. 3 of Grindlay et al (2009; from Salvaterra 2009) for the comparative sensitivities and energy band coverage and same assumed SFR vs. *z*. From photometric redshift constraints, a second *Swift* GRB at *z* ~ 9-10 (GRB090429B) is likely (Tanvir 2010, these Proceedings). Given that 280 *Swift* GRBs now have optical/nIR IDs, this 2/280 = 0.7% fraction is consistent with Fig. 3 and suggests *EXIST* would measure redshifts and spectra for ≥30 GRBs/year at *z* > 8. Since the Ly-α DLA sensitivity to the IGM ionization fraction decreases for host galaxies with internal Hydrogen columns NH ≥ $10^{20}$cm$^{-2}$ (though at z >7, lower values might be expected in the ~$10^9$ $M_\odot$ hosts with low metallicity, as also suggested by the limits found for GRB090423), then ≥30-100 EoR measures at z >8 are expected over a 5y mission. These in situ measures will constrain the EoR vs. z and spatial scales of patchiness of the ionized IGM vs z.

## Comparison with LAE galaxy and photo-z measures of the EoR

Ly α emission (LAE) galaxies, as the likely agents of reionization, have long been suggested as probes of the EoR (e.g. Malhotra and Rhoads 2006 and references therein). Since the AGN number density is rapidly falling at z >6 (Fan et al 2001), it is now believed that star forming dwarf galaxies, which would be LAEs, are the dominant sources of reionization. How might the current and projected LAE surveys (e.g. Tilvi et al 2010) for LAEs compare with GRBs as tracers of the EoR? A full answer is beyond the scope of this paper, but in short: LAEs are difficult to find (as are, currently, high-z GRBs!). The problems are multi-fold, but primarily that they are (as dwarfs) faint and with (only) a “narrow-band” signature (the LAE itself), with typical rest frame detection limits of EW(Ly α) >5-10Angstroms. The big problem, though, is the “OH Forest” or high background of night sky emission that plagues all ground-based surveys, as shown in Fig. 3, in which the background at two resolutions is plotted vs. redshift z. The LAE surveys must be done in the narrow windows (e.g. the three principal R = 300 bands shown). Interestingly, the first GRB to have its redshift measured at z >7 is nearly centered in one of these "windows". While probably at least in part a coincidence, it does point out the bias that must be present in detections of the Ly $\alpha$ break from ground-based spectra or photo-z's: GRBs well off these windows are more likely to appear "dark". The LAE surveys benefit of course from non-fading sources but are limited by the restricted field of view of narrow-band filters on large telescope optics. Several LAE candidates have been found at z ≤ 7.7 (Tilvi et al 2010 and references therein), but it is not clear that the technique can be pushed beyond the next windows at z ~8.8 and ~10.1.

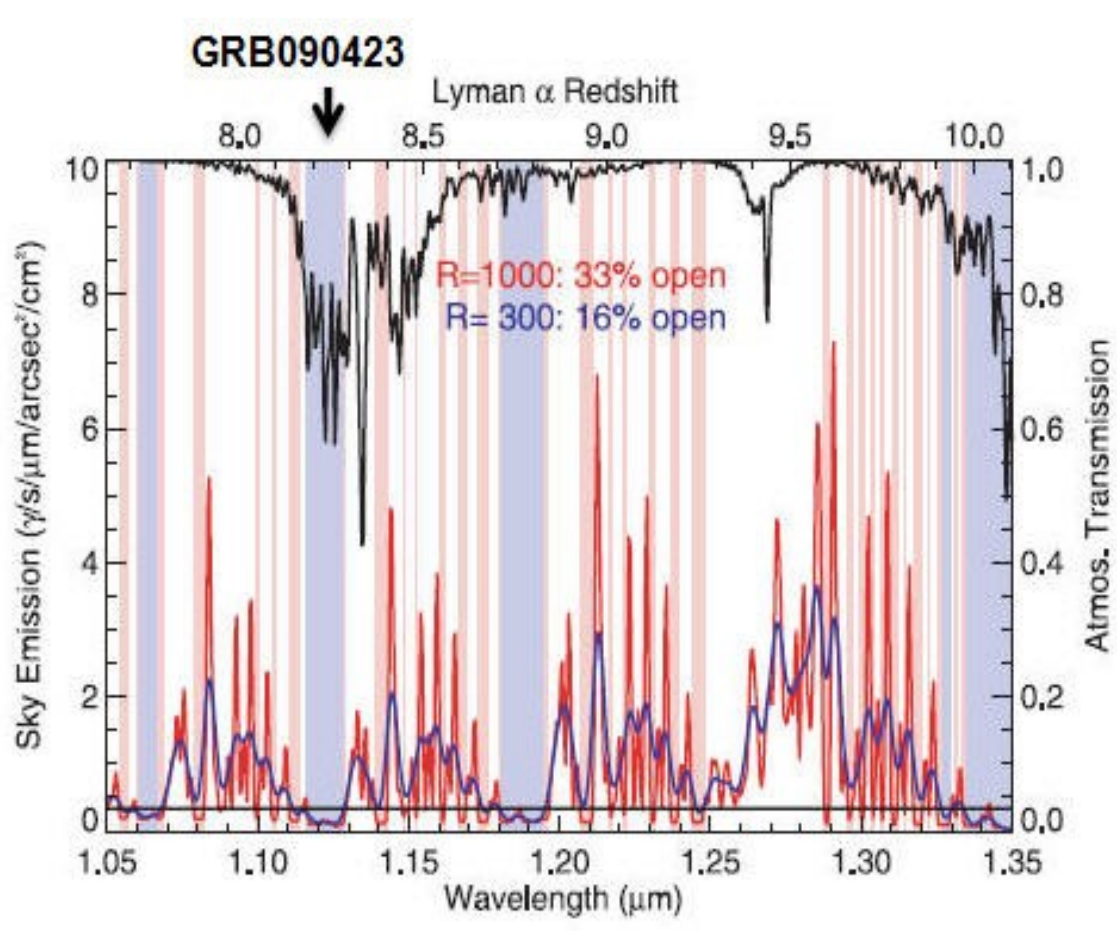

**FIGURE 3.** OH Forest vs. z (Barton et al 2004).

Searches for Y or (even) J band dropouts with HST/WFC3 with sensitivities reaching AB ~29 have now revealed 66 z ~7 and 47 z ~8 candidate galaxies (Bouwens et al 2010 and references therein) which of course now constitute the largest sample of galaxy candidates in the EoR. These exciting results already provide constraints on the ionization fraction vs. ionizing photon escape fractions at z ~6 – 8 and appear consistent with the WMAP values for Thomson optical depths which indicate ~50% reionization by z ~10.5. Nevertheless, there is a significant role for future GRB observations, which given the initial (and for some hours) magnitudes that are ≥10 magnitudes brighter than the dwarf galaxy hosts, allow the best hope for high resolution spectroscopy to measure ionization fractions and metallicities in detail at these early epochs. Since at z > 8 even JWST cannot resolve the "typical" host galaxies, it may only be from moderate to high resolution spectra of GRB afterglows at magnitudes reaching AB ~24 that the structure of the first galaxies can be "resolved". Salvaterra et al (2010) have shown how PopIII stars are unlikely to be detectable given their <5% contribution to the total luminosity of early galaxies. If PopIII stars either above or below the Pair Instability Supernova (PISN) limit collapse to produce GRBs, at any redshift (survivors could exist up to z ~6?), then their expected higher luminosity and longer duration (e.g. Meszaros and Rees 2010) will be readily detectable (though rare). Again, only a prompt high resolution nIR spectrum (from space) could distinguish the very different composition (e.g. strong HeII $\lambda$1640, detectable with the IRT up to z ~ 12.5) of the ISM in such a system.

## GRBs as probes of SFR(z) and Z(z)

The redshift evolution of the cosmic star formation rate, SFR(z), and metallicity, Z(z), are both key objectives for *EXIST*. Even for heavily absorbed ("dark") GRBs, the great nIR sensitivity will allow redshifts for most GRBs at z <6. Thus a very high fraction of GRBs at z <5-6 will likely have their redshifts measured in a "uniform" way, providing by far the largest measure of SFR(z) for the ≥20 $M_\odot$ stars thought to be the progenitors of Long GRBs (LGRBs). Present studies of GRB hosts and their mass and metallicities indicate that LGRB hosts are predominantly dwarf star forming galaxies (Savaglio et al 2009). This is "convenient" in that these are the galaxies expected and now observed (Bouwens et al) at z ~8 (and probably up to z ≥10). Thus the SFR(z) derived from simply measuring the rate dN(GRBs)/dz for GRBs above well defined detection (and redshift/spectral followup) thresholds will be well defined.

Each of the ~3000 measures expected will also have high resolution (0.15”) 4-band imaging) so that some constraints on morphological type (for low-z) and local environment (galaxy groups would be recognized in objective prism spectra of the field obtained “free” during the GRB integration). Finally, from the R = 3000 spectra expected for a significant fraction (≥50-70%) of the total, metallicities will be derived. While not high resolution enough for detailed [Fe/H] measures, they will in a uniform way trace the broad dependence of Z(z). The high quality SXI soft X-ray spectra for every GRB will yield abundance measures for CNO, S, Mg, Fe from absorption edge measures which also allow, for bright bursts, redshifts directly (Campana et al 2010).

## Long vs. Short GRBs and Short GRBs as standard candles?

The high sensitivity and broader bandwidth will allow definitive studies of GRB physics with high temporal resolution spectra and polarization. From its 3D event position detection capability (Hong et al 2010), the HET enables “easy” polarization measures for GRBs and Blazars. Depending on upcoming small mission polarization studies (e.g. the recently launched IKAROS), this capability could be enhanced. High time resolution spectra of GRBs (and bright sources) will be a key resource to constrain models, finally, of the central engine. For example, the still open question of whether GRB jets are powered by an electromagnetic Poynting flux vs. an expanding fireball (see Gehrels et al 2010 for a review) could be settled by both high temporal and spectral resolution and polarization studies. The physics of these most energetic events since the Big Bang can surely be understood with the multi-wavelength, high time resolution spectroscopic capabilities proposed for *EXIST*.

Although the short GRBs (SGRBs) are clearly a separate and lower luminosity population than LGRBs and likely due to binary mergers, *EXIST* will constrain the nature (NS-NS vs. NS-BH mergers?) and host galaxy population. The offset postitions from host galaxies point to either binary ejection or, more likely (since NS-NS pulsar binaries in the Milky Way do *not have high proper motion*), products of exchange collisions of isolated NSs with NS-WD/ms binaries in globular clusters (Grindlay et al 2006). The NS-NS mergers will be readily detectable as gravitational wave in-spirals with Advanced LIGO if within ~200 Mpc, for which the well measured IRT positions and spectra from *EXIST* will enable precision Hubble constant measurements (Bloom et al 2009). For those with IRT detections offset from galaxies, the 0.1” positions will enable JWST to search for globular cluster hosts, with AB ~30 for a SGRB/GC at z = 0.1, and offsets ~5” for a GC at 10 kpc from its host.

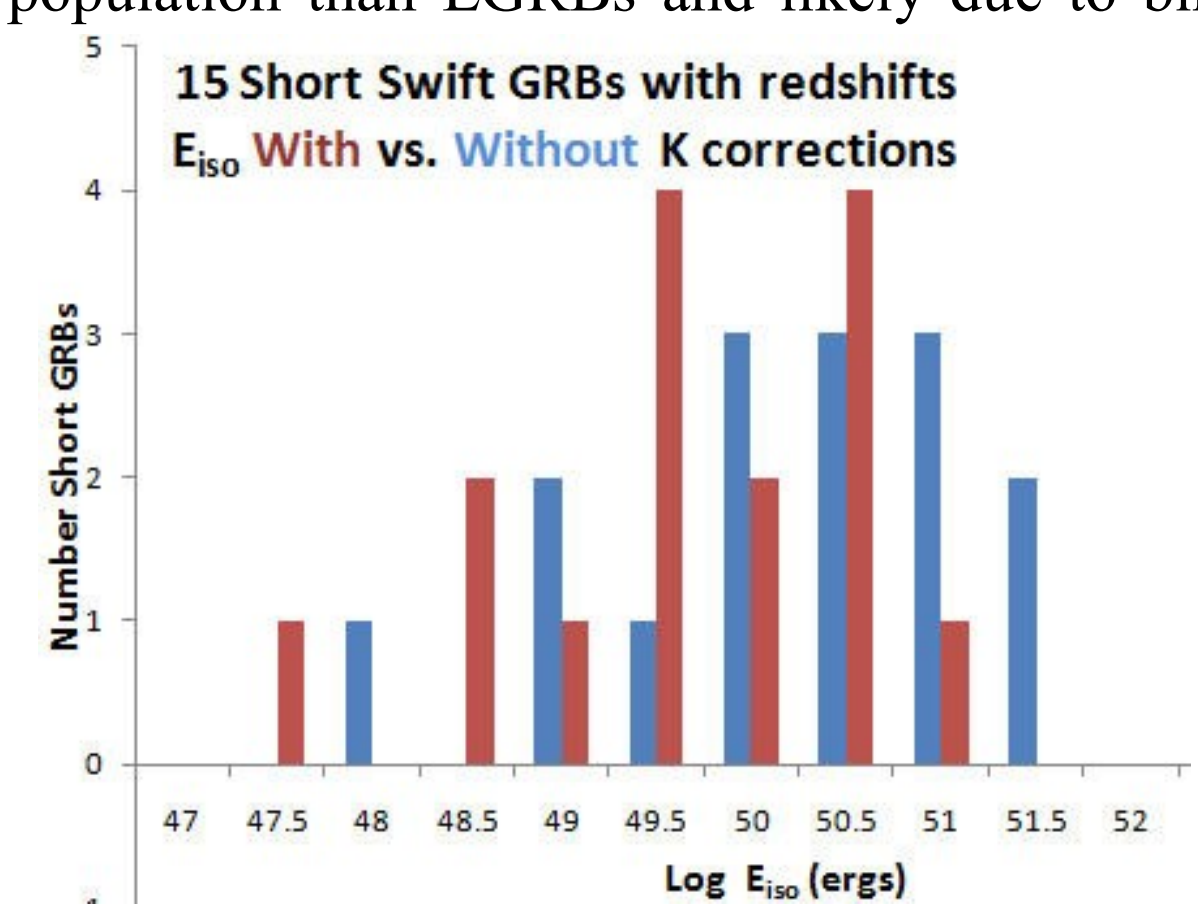

**FUGURE 4.** $E_{iso}$ distributions for all 15 SGRBs with redshifts. Larger samples with broader band spectra are needed to test whether K-corrected distributions could be used as standard candles.

Finally, if SGRBs are predominantly NS-NS mergers, and given the small mass range of most dynamically measured NSs (1.4 ±0.2 $M_\odot$), they might be expected to have relatively constant isotropic equivalent energy release, $E_{iso}$. Since beaming appears to be less for SGRBs than for LGRBs, beaming corrections may broaden the $E_{iso}$ distribution less than for LGRBs. Accordingly, we have extracted the $E_{iso}$ distribution for the 15 SGRBs with reasonably secure redshifts using just the observed fluence in the 15-150 keV BAT band. We then redshift the fluence and allow for the measured BAT PL spectral shape (well determined, but only over this band) to derive $E_{iso}$ in the source frame. The resulting distributions are shown in Fig. 4. The results are suggestive but clearly a much larger sample, with broader band spectra, are needed to determin if SGRBs could conceivably be "standard candles" out to z ~2-3 where they should still be detectable by *EXIST*. If so, they might extend/complement SNIas.

## READY TO *EXIST*

Given the successful balloon flight test (October, 2009) of *ProtoEXIST1* employing key technologies for HET (Hong et al 2010) and the current development of finer pixel CZT imagers derived from *NuSTAR*, and with the flight qualified key components for both IRT (Kutyrev et al 2009) and SXI (Tagliaferri et al 2009), *EXIST* could be launched (by an ESA Soyuz, at reduced cost, and into a low background orbit) in 2017. We are grateful for support from NASA ASMC grant NNX08AK84G.